\documentclass{phbauth}        % Don't modify
\usepackage{graphicx}

\begin{document}

\begin{frontmatter}

% Use lower case letters in the title.
\title{Superconducting Gap and Pseudogap in Bi--2212}

\author[address1]{Matthias Opel\thanksref{thank1}},
\author[address1]{Francesca Venturini},
\author[address1]{Rudi Hackl},
\author[address2]{Helmuth Berger},
\author[address2]{L\'aszl\'o Forr\'o}

\address[address1]{Walther--Meissner--Institut,
Bayerische Akademie der Wissenschaften, 85748 Garching, Germany}
\address[address2]{IGA, Ecole Polytechnique F\'ed\'erale, 1015 Lausanne, Switzerland}

% The corresponding author should be distinguished and his email
% address and/or fax number must be given. His mailing address has to
% be complete: the proofs are send to this address around
% January 1, 2000. The address for sending proofs has to be indicated
% as "present address", if it is different from the address above.

\thanks[thank1]{Corresponding author. E-mail: opel@badw.de} 

\begin{abstract}

We present results of Raman scattering experiments in differently doped
Bi$_{2}$Sr$_{2}$(Ca$_{x}$Y$_{1-x}$)Cu$_{2}$O$_{8+\delta}$ (Bi--2212)
single crystals.
Below $T_c$ the spectra show pair--breaking features in the
whole doping range reflecting a $d_{x^2-y^2}$ order parameter.
In the normal state between $T_c$ and $T^\ast \simeq 200 {\rm K}$
we find evidence for a pseudogap in B$_{2g}$ symmetry.
Upon doping its effect on the spectra decreases while its energy scale
appears to be unchanged.

\end{abstract}


\end{frontmatter}

% The main text begins here. The \section commands are optional.

\begin{figure*}[t]
%h=here, t=top, b=bottom, p=separate figure page
\begin{center}
\leavevmode
\includegraphics[width=13cm]{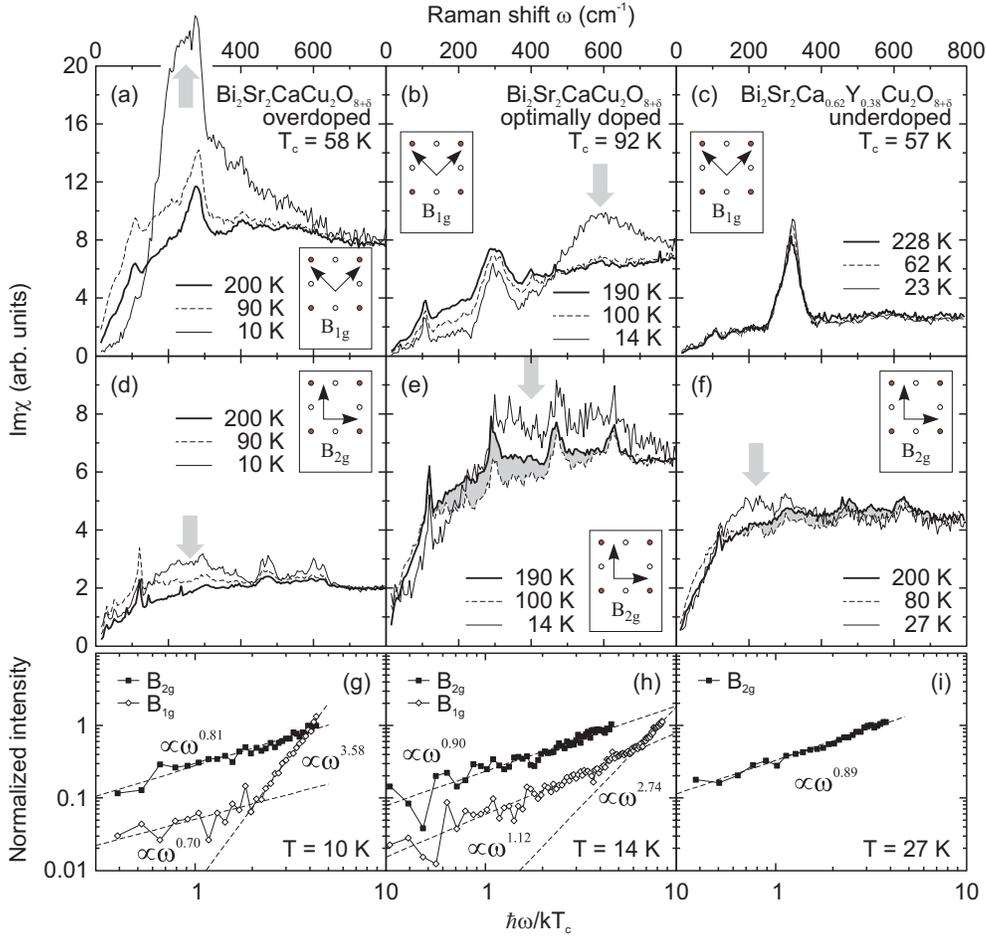}
\caption{Raman response ${\rm Im}\chi(\omega,T)$ in B$_{1g}$ (a-c) and B$_{2g}$
symmetry (d-f) for Bi--2212. The pair-breaking peaks below $T_c$ are marked with
vertical arrows. For the low frequency behavior in the SC state (g-i)
the spectra (with phonons subtracted)
have been normalized to 1 at $800 {\rm cm}^{-1}$. The dashed lines are linear fits
to the data on logarithmic scale functions, the obtained exponents are indicated.
The shaded areas in (e,f) show the loss of spectral weight in the PG state.}
\label{Bild}
\end{center}
\end{figure*}

The relationship between the superconducting (SC) and the pseudogap (PG) phases
and their evolution with doping is of particular interest for understanding
the cuprates. In the following we will describe
recent results from light scattering experiments in
Bi$_{2}$Sr$_{2}$(Ca$_{x}$Y$_{1- x }$)Cu$_{2}$O$_{8+\delta}$
(Bi--2212) single crystals.
The experiments were performed in pseudo back-scattering geometry
using a standard Raman set-up.

Figure~\ref{Bild} shows spectra from Bi--2212 at three different doping levels
as a function of temperature. The raw data have been divided by the Bose factor
in order to obtain the Raman response ${\rm Im}\chi(\omega,T)$.
In the SC state the well-known pair-breaking peaks (marked with vertical arrows)
can be found in both B$_{1g}$ (a-c) and B$_{2g}$ (d-f) symmetry
for high and optimal doping whereas in the underdoped material this feature
is only present in B$_{2g}$ symmetry. The peak frequencies in the B$_{2g}$ channel
and thus the energy scales of the SC correlations
depend on the doping level and scale with the transition temperature $T_c$.
In the B$_{1g}$ channel the intensity
of the pair-breaking peak decreases continuously when approaching the underdoped range
of the phase diagram which goes along with a strong suppression of the overall Raman intensity.
This effect may be caused by a change of the Fermi surface~(FS) topology
and is also seen in YBa$_{2}$Cu$_{3}$O$_{6+x}$~\cite{Chen}
and La$_{2-x}$Sr$_x$CuO$_4$~\cite{Irwin}. 
The low frequency behavior of the SC spectra is shown in Fig.~\ref{Bild}~(g-i)
in a double logarithmic representation. In the B$_{2g}$ channel the intensities increase
almost linearly over a large frequency range while in B$_{1g}$ we find a cross-over
from linear to cubic behavior. These characteristic power laws 
are strong evidence for $d_{x^2-y^2}$ wave pairing \cite{DevEin}
in the whole doping range.

In the normal state at temperatures between $T^\ast \simeq 200 {\rm K}$ 
and $T_c$ the B$_{2g}$ response shows new features: Spectral weight is lost within an 
intermediate, doping-independent frequency interval (shaded areas in Fig.~\ref{Bild}~(e,f)).
This effect is a signature for the opening of a PG $\Delta^\ast$ below $T^\ast$
in the quasiparticle excitation spectrum \cite{unserPG} which disappears
at higher doping levels (Fig.~\ref{Bild}~(d)).
Due to details of the spectral shape we conclude that $\Delta^\ast$
as well as the SC order parameter $\Delta$
have d-type topology. However, the gapped region in the PG state
around $(\pi,0)$ must be quite narrow.
Such a scenario has been modeled by confining the FS integration to the vicinity
$(\pi/2,\pi/2)$. Then, the B$_{2g}$ response is almost unchanged in the underdoped regime 
while in B$_{1g}$ symmetry most of the pair-breaking peak is cut off \cite{Chen,Irwin}.

% Acknowledgements are optional.
\begin{ack}

We benefitted a lot from the continuous collaboration with Tom Devereaux and
Dietrich Einzel.

\end{ack}

% References

\end{document}